\documentclass[8.5pt,twoside,twocolumn]{article}
\usepackage[super,sort&compress,comma]{natbib} 
\usepackage[version=3]{mhchem}
\usepackage{balance}
\usepackage{times,mathptm}
\usepackage{sectsty}
\usepackage{graphicx}
\usepackage{lastpage}
\usepackage[format=plain,justification=raggedright,singlelinecheck=false,font=small,labelfont=bf,labelsep=space]{caption} 
\usepackage{fancyhdr}
\graphicspath{{fig/}}
\begin{document}
\thispagestyle{plain}
\fancypagestyle{plain}{
\fancyhead[L]{\includegraphics[height=8pt]{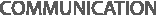}}
\fancyhead[C]{\hspace{-1cm}\includegraphics[height=20pt]{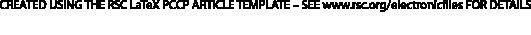}}
\fancyhead[R]{\hspace{10cm}\vspace{-0.25cm}\includegraphics[height=10pt]{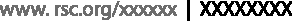}}
\renewcommand{\headrulewidth}{1pt}}
\renewcommand{\thefootnote}{\fnsymbol{footnote}}
\renewcommand\footnoterule{\vspace*{1pt}%
\hrule width 3.4in height 0.4pt \vspace*{5pt}} 
\setcounter{secnumdepth}{5}
\makeatletter 
\renewcommand\@biblabel[1]{#1}            
\renewcommand\@makefntext[1]%
{\noindent\makebox[0pt][r]{\@thefnmark\,}#1}
\makeatother 
\renewcommand{\figurename}{\small{Fig.}~}
\sectionfont{\large}
\subsectionfont{\normalsize} 
\fancyfoot{}
\fancyfoot[LO,RE]{\vspace{-7pt}\includegraphics[height=9pt]{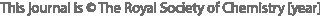}}
\fancyfoot[CO]{\vspace{-7.2pt}\hspace{12.2cm}\includegraphics{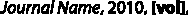}}
\fancyfoot[CE]{\vspace{-7.5pt}\hspace{-13.5cm}\includegraphics{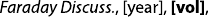}}
\fancyfoot[RO]{\footnotesize{\sffamily{1--\pageref{LastPage} ~\textbar  \hspace{2pt}\thepage}}}
\fancyfoot[LE]{\footnotesize{\sffamily{\thepage~\textbar\hspace{3.45cm} 1--\pageref{LastPage}}}}
\fancyhead{}
\renewcommand{\headrulewidth}{1pt} 
\renewcommand{\footrulewidth}{1pt}
\setlength{\arrayrulewidth}{1pt}
\setlength{\columnsep}{6.5mm}
\setlength\bibsep{1pt}
\twocolumn[
\begin{@twocolumnfalse}
\noindent\LARGE{\textbf{Master equation for the probability distribution functions of forces in soft particle packings$^\dag$}}\vspace{0.6cm}
\noindent\large{\textbf{Kuniyasu Saitoh,$^{\ast}$ Vanessa Magnanimo, and Stefan Luding}}\vspace{0.5cm}
\noindent\textit{\small{\textbf{Received Xth XXXXXXXXXX 20XX, Accepted Xth XXXXXXXXX 20XX\newline
First published on the web Xth XXXXXXXXXX 200X}}}
\noindent\textbf{\small{DOI: 10.1039/b000000x}}\end{@twocolumnfalse}\vspace{0.6cm}
]
\noindent\textbf{
Employing molecular dynamics simulations of jammed soft particles,
we study microscopic responses of force-chain networks to quasi-static isotropic (de)compressions.
We show that not only contacts but also interparticle gaps between the nearest neighbors must be considered
for the stochastic evolution of the probability distribution functions (PDFs) of forces,
where the mutual exchange of contacts and interparticle gaps,\ i.e.\ opening and closing contacts, are also crucial to the incremental system behaviors.
By numerically determining the transition rates for all changes of contacts and gaps,
we formulate a Master equation for the PDFs of forces,
where the insight one gets from the transition rates is striking:
The mean change of forces reflects non-affine system response, while their fluctuations obey uncorrelated Gaussian statistics.
In contrast, interparticle gaps are reacting mostly affine in average, but imply multi-scale correlations according to a wider stable distribution function.
}
\section*{}
\vspace{-1cm}
\footnotetext{\dag~Electronic Supplementary Information (ESI) available:
[details of any supplementary information available should be included here].
See DOI: 10.1039/b000000x/}
%
\footnotetext{\textit{$^{a}$~Faculty of Engineering Technology, MESA+, University of Twente, Drienerlolaan 5, 7522 NB, Enschede, The Netherlands}}
%
%
Quasi-static deformations of soft particles,\ e.g.\ glasses, colloids, emulsions, foams, and granular materials,
have been widely investigated because of their significant importance in industry and science.
However, many challenges of describing their macroscopic behaviors still remain due to disordered configurations, complex dynamics, etc \cite{lemaitre}.
At the microscopic scale, mechanical responses of soft particle packings are probed as a reconstruction of force-chain networks \cite{an0,gn2},
where complicated non-affine displacements of particles cause the ``recombination" of force-chains,\ i.e.\ opening and closing contacts \cite{merlijn}.
Once a macroscopic quantity is defined as a statistical average in force-chains,\ e.g.\ the stress tensor, elastic moduli, etc,
its non-trivial response to quasi-static deformations (i.e.\ \emph{non-affine response}) is governed by the change of the probability distribution function (PDF) of forces.
Therefore, the PDFs in soft particle packings have practical importance so that a lot of theoretical studies
(e.g.\ based on the stress ensemble \cite{ft4}, force network ensemble \cite{en4}, entropy maximization \cite{en7}, and so on \cite{ft0,ft3})
have been devoted to determine their functional forms observed in experiments \cite{ps1,ps2} and numerical simulations \cite{pd3,pd4}.
In general, the PDFs are \emph{asymmetric} and cannot be described by conventional distribution functions \cite{gu4}.
Moreover, there is still much debate about their tails \cite{pd0,gu5} as well as their shapes for small forces \cite{wyart0,wyart1,wyart2}.

\begin{figure}[h]
\centering
\includegraphics[width=8.5cm]{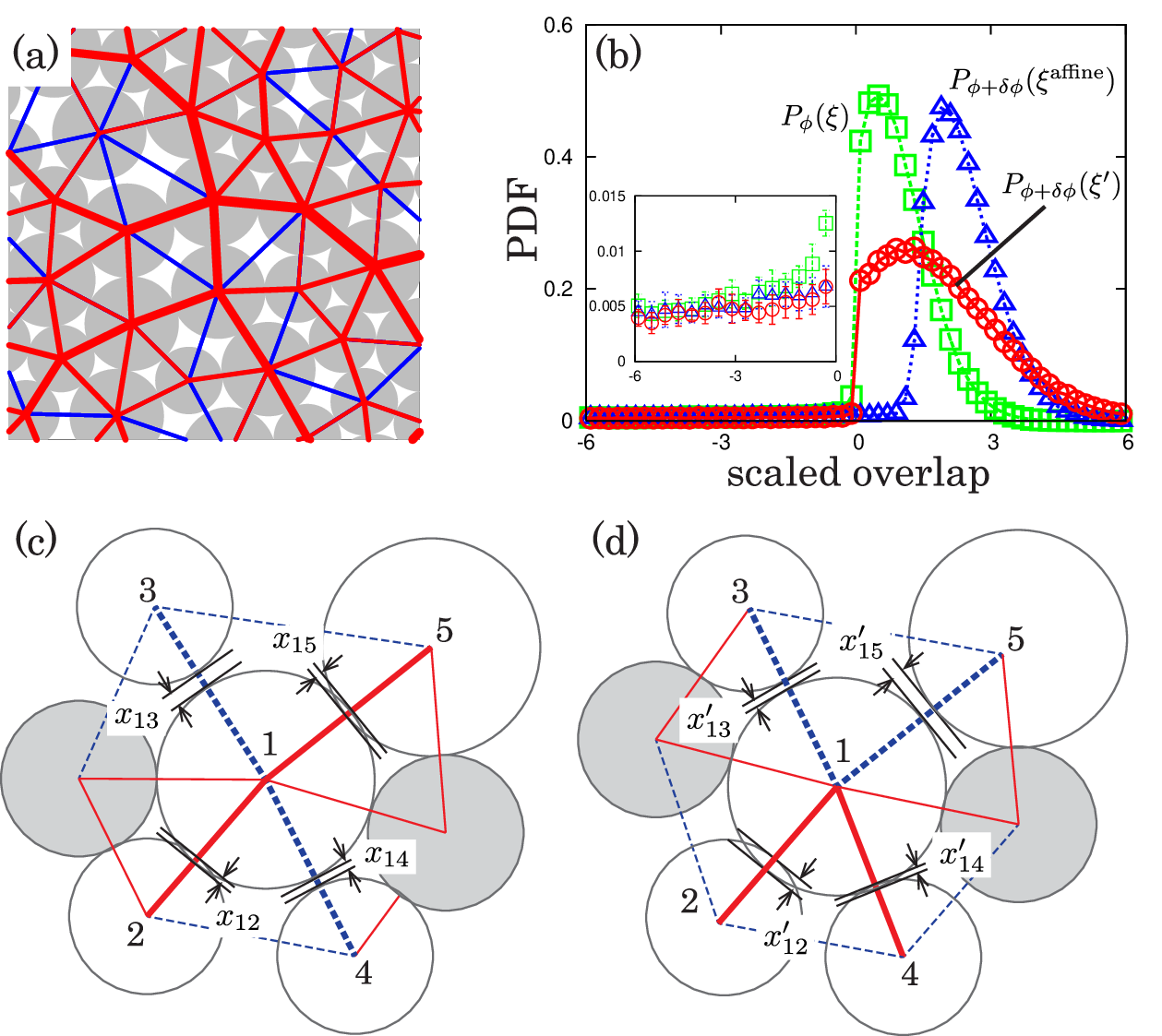}
\caption{(Color online)
(a) Sketch of the generalized force-chain network with contacts (red lines) and virtual contacts (blue lines),
where overlaps are defined as positive and negative, respectively.
The widths of red lines are proportional to the strength of forces.
(b) The PDFs of scaled overlaps, $P_\phi(\xi)$ (squares), $P_{\phi+\delta\phi}(\xi^\mathrm{affine})$ (triangles),
and $P_{\phi+\delta\phi}(\xi')$ (circles), for $\phi-\phi_J=1.2\times10^{-3}$ and $\delta\phi=1.2\times10^{-3}$.
The inset is the zoom-in to the PDFs of virtual contacts.
(c) and (d): Sketches of the DT around a single particle (c) before compression and (d) after relaxation,
where red solid and blue dashed lines represent contacts and virtual contacts, respectively.
The circles are particles with centers placed on the Delaunay vertices.
\label{fig:delaunay}}
\end{figure}

In this study, we propose a new method for describing the evolution of the PDFs of forces under quasi-static deformations.
Employing the Delaunay triangulation (DT) for two-dimensional packings (see Fig.\ \ref{fig:delaunay}(a)),
we generalize the ``overlap" between particles ($i$ and $j$) connected by a Delaunay edge as
\begin{equation}
x_{ij} \equiv R_i+R_j-D_{ij}~,
\label{eq:overlap}
\end{equation}
where $R_i+R_j$ and $D_{ij}$ are the sum of radii and the Delaunay edge length, respectively,
so that not only \emph{contacts} ($x_{ij}>0$), but also interparticle gaps or \emph{virtual contacts} ($x_{ij}<0$) can be included in force-chain networks
\footnote[3]{
Since the DT is unique for each packing, virtual contacts are uniquely determined,
where the total number of contacts and virtual contacts is a conserved quantity which is independent of the area fraction.
We have not observed any \emph{flips} of the Delaunay edges if $\gamma\le10^{-3}$,
and the number of flipped edges are less than $1\%$ at most for $\gamma\sim10$.}.
We then apply quasi-static isotropic (de)compressions to the packings, where the area fraction, $\phi$, increases (or decreases) by $\delta\phi$
and the PDF of generalized overlaps,\ Eq.\ (\ref{eq:overlap}), captures the statistics of contacts and virtual contacts after opening or closing contacts.
Our main result is that we numerically calibrate a Master equation for the PDFs of generalized overlaps,
where transition rates of generalized overlaps are \emph{symmetric} and can be described by conventional distribution functions.
In addition, we find that the transition rates depend on both an applied strain step, $\delta\phi$, and the distance from jamming point, $\phi-\phi_J$,
through only one scaling parameter, $\gamma\equiv\delta\phi/(\phi-\phi_J)$, where $\phi_J$ is the area fraction at jamming.
The Master equation is able to describe all features of the PDFs,\ e.g.\ their changes during compressions
and discontinuous ``jumps",\ i.e.\ restructuring around zero-overlaps, which had been observed in a previous study \cite{th1}.
The application perspective of our method is that it allows us to compute the local energy density given by the second moment of particle overlaps
as a statistical approach to large scale problems.
The hydrostatic pressure and bulk modulus can be deduced from the first and second derivatives of the energy density, respectively,
where the derivatives are defined by the Master equation (see the ESI \dag).

As method, we use molecular dynamics (MD) simulations of two-dimensional frictionless soft particles.
The normal force between particles in contact ($i$ and $j$) is given by $f_{ij}=kx_{ij}-\eta\dot{x}_{ij}$~($x_{ij}>0$)
with a spring constant, $k$, viscosity coefficient, $\eta$, and relative speed in the normal direction, $\dot{x}_{ij}$.
A global damping force, $\mathbf{f}^\mathrm{d}_i=-\eta\mathbf{v}_i$, proportional to the particle's velocity, $\mathbf{v}_i$,
is also introduced to enhance the relaxation, where the particles lose their kinetic energy by means of inelastic contacts and global damping.
We randomly distribute a $50:50$ binary mixture of $N$ particles with two kinds of radii, $R_i>R_j$ ($R_i/R_j=1.4$), in a square periodic box,
where no particle touches others.
We then rescale every radius to make mechanically stable particle packings (our method is similar to the one used in Ref.\ \cite{rs1}
\footnote[4]{
We rescale every radius as $R(t+\delta t)=[1+\{\bar{x}-x_\mathrm{m}(t)\}/l]R(t)$, where $t$, $\delta t$, $\bar{x}$, and $x_\mathrm{m}(t)$
are time, increment of time, target mean overlap, and averaged overlap at time $t$, respectively.
When $\bar{x}>x_\mathrm{m}(t)$, each radius increases, while it decreases if $\bar{x}<x_\mathrm{m}(t)$.
Therefore, the averaged overlap converges to the target value, $\bar{x}$, in the long time limit.
Here, we keep the mass constant and use a long length scale $l=10^2\bar{R}$ to grow the particles gently, where $\bar{R}$ is the mean radius at $t=0$.
Note that the static packings prepared with longer length scales, $l=10^3\bar{R}$ and $10^4\bar{R}$, give the same results
concerning critical scaling of frictionless particles near jamming \cite{gn3}, while we do not obtain the same results with $l=10\bar{R}$.
We stop rescaling each radius when every acceleration of particles drops below a threshold $10^{-6}k\bar{R}/m$ and assume the system is static.
}).
In our simulations, distances from jamming are determined by the known scaling of averaged overlap \cite{gn3,sastry}, $\bar{x}(\phi)\simeq A(\phi-\phi_J)$.
From our $10$ samples of $N=8192$ particles, we estimate $\phi_J=0.8458\pm10^{-4}$ with a critical amplitude, $A=(0.31\pm0.01)\bar{\sigma}$,
where $\bar{\sigma}$ is the mean diameter in a packing closest to the jamming point, $\phi-\phi_J=1.2\times10^{-5}$.
We also prepared $10$ samples for small systems ($N=512,2048$) and $2$ samples for the largest one ($N=32768$),
while we only report the results of $N=8192$ since none of the results depends on system size (see the ESI \dag).

We apply an isotropic compression to the packings by multiplying every radius by $\sqrt{1+\delta\phi/\phi}$,
where the area fraction increases from $\phi$ to $\phi+\delta\phi$.
At the same time, all the generalized overlaps, $x_{ij}$, change to $x_{ij}^\mathrm{affine}=x_{ij}+(D_{ij}/2\phi)\delta\phi$
\footnote[5]{We neglected the higher order term proportional to $x_{ij}\delta\phi$}.
However, the particles are randomly arranged and their force balance is broken by compression so that the system is allowed to relax to a new mechanically stable state
\footnote[6]{
From our results of the mean square displacements, most particles do not jump out of cages and our systems do not undergo structural relaxations after compression.
We also checked that the response to compression does not depend on the protocols,\ e.g.\ an overdamped dynamics.
}.
After relaxation, the overlaps change to new values, $x'_{ij}\neq x_{ij}^\mathrm{affine}$,
due to non-affine displacements of the particles, where we observe four kinds of changes (from $x_{ij}$ to $x'_{ij}$) as shown in Figs.\ \ref{fig:delaunay}(c) and (d):
$x_{12}>0$ and $x_{13}<0$ change to $x'_{12}>0$ and $x'_{13}<0$, respectively, where they do not change their signs and thus contacts are neither generated nor broken.
We name these changes ``contact-to-contact (CC)" and ``virtual-to-virtual (VV)", respectively.
On the other hand, $x_{14}<0$ and $x_{15}>0$ change to $x'_{14}>0$ and $x'_{15}<0$, respectively, where a new contact is generated and an existing contact is broken, respectively.
We call these changes ``virtual-to-contact (VC)" and ``contact-to-virtual (CV)", respectively.

The restructuring of the force-chains, attributed to the changes, (CC), (VV), (VC), and (CV), is well captured by the PDFs of the generalized overlaps.
Figure \ref{fig:delaunay}(b) displays the PDFs of the overlaps scaled by the averaged overlap before compression,
$\xi\equiv x_{ij}/\bar{x}(\phi)$, $\xi^\mathrm{affine}\equiv x_{ij}^\mathrm{affine}/\bar{x}(\phi)$, and $\xi'\equiv x'_{ij}/\bar{x}(\phi)$,
where we omit the subscript $ij$ from the scaled overlaps.
As can be seen, the difference between affine and non-affine deformations is clear:
The affine deformation just shifts the PDF before compression to the positive direction,
while non-affine deformations broaden the PDF in positive overlaps and reconstruct the discontinuous ``jump" around zero.
Note that, however, the new PDF in negative overlaps is comparable with that after affine deformation (see the inset in Fig.\ \ref{fig:delaunay}(b)).

To describe such non-affine evolution of the PDFs, we introduce the Chapman-Kolmogorov equation \cite{vanKampen},
\begin{equation}
P_{\phi+\delta\phi}(\xi') = \int_{-\infty}^\infty W(\xi'|\xi)P_\phi(\xi)d\xi~,
\label{eq:conditional}
\end{equation}
where $W(\xi'|\xi)$ is a conditional probability distribution (CPD) satisfying the normalization condition, $\int_{-\infty}^\infty W(\xi'|\xi)d\xi'=1$.
The CPD is the probability of overlaps becoming $\xi'$ which were $\xi$ before compression (i.e.\ a distribution of $\xi'$ around a mean value which depends on $\xi$).
For example, the CPD for affine deformation is a delta function, $W_\mathrm{affine}(\xi'|\xi)=\delta(\xi'-f_a(\xi))$,
where the mean value is given by a linear function of $\xi$, $f_a(\xi)=\xi+B_a\gamma$, with a coefficient, $B_a=D_{ij}/(2A\phi)$,
which just shifts the PDF by $B_a\gamma$,\ i.e.\ $P_{\phi+\delta\phi}(\xi)=P_\phi(\xi-B_a\gamma)$, as shown in Fig.\ \ref{fig:delaunay}(b)
\footnote[7]{$\gamma$ can be large, whereas $\delta\gamma$ is always small.}.

\begin{figure}[h]
\centering
\includegraphics[width=10.5cm]{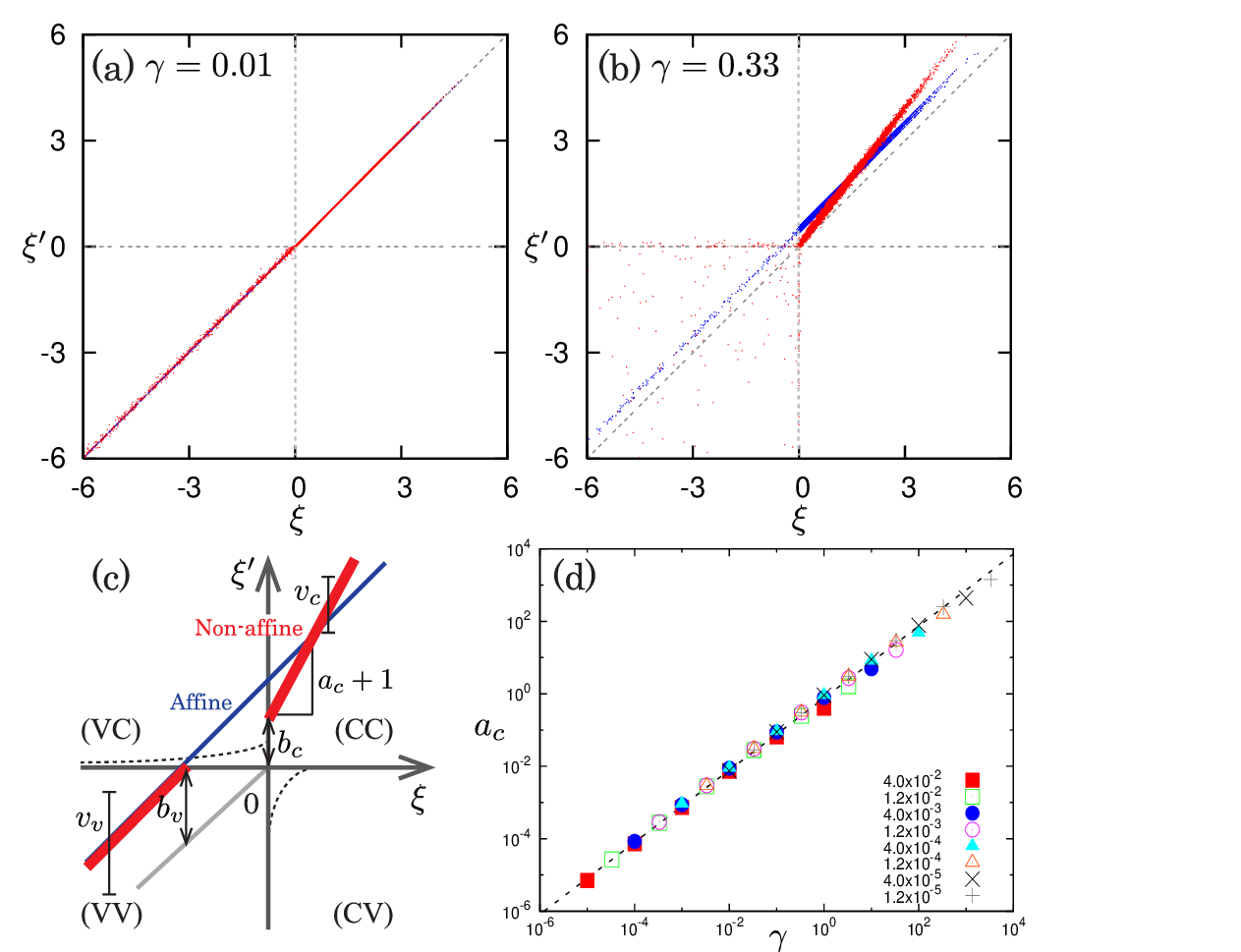}
\caption{(Color online)
(a) and (b): Scatter plots of overlaps, where the blue and red dots are affine and non-affine responses to compression, $(\xi,\xi^\mathrm{affine})$ and $(\xi,\xi')$, respectively.
Here, $\delta\phi=4\times10^{-5}$ and $\phi-\phi_J=$ (a) $4\times10^{-3}$ ($\gamma=0.01$) and (b) $1.2\times10^{-4}$ ($\gamma=0.33$).
(c) A sketch of deviations from an affine deformation, where the blue and red solid lines represent $f_a(\xi)$ (for small and large particles) and $f_n(\xi)$ $(n=c,v)$, respectively.
(d) A double logarithmic plot of $a_c$ against $\gamma$, where $\delta\phi$ is ranged between $4\times10^{-7}\le\delta\phi\le4\times10^{-3}$,
and different symbols represent different distances from jamming, $\phi-\phi_J$, as given in the inset.
\label{fig:scat}}
\end{figure}

On the other hand, the CPDs for non-affine deformations can be measured through scatter plots of the scaled overlaps,
see Figures \ref{fig:scat}(a) and (b), where the four kinds of changes are mapped onto four regions:
(CC) $\xi,\xi'>0$, (VV) $\xi,\xi'<0$, (VC) $\xi<0$, $\xi'>0$, and (CV) $\xi>0$, $\xi'<0$, respectively.
In (CC) and (VV), the scaled overlaps after compression distribute around mean values which we describe by linear fitting functions for $\xi'$,
\begin{equation}
f_n(\xi) = (a_n+1)\xi+b_n~,
\label{eq:mzt}
\end{equation}
where the subscripts, $n=c$ and $v$, represent the mean values in (CC) and (VV), respectively.
If we introduce standard deviations of $\xi'$ from $f_n(\xi)$ as $v_n$, which are almost independent of $\xi$,
the systematic deviation from affine deformations can be quantified by the coefficients, $a_n$, $b_n$, and $v_n$, as summarized in Fig. \ref{fig:scat}(c).
Note that the differences are always present, but not visible if the applied strain is small or the system is far from jamming,\ i.e.\ if $\gamma\ll1$ (Fig.\ \ref{fig:scat}(a)),
while $\xi'$ deviates more from $f_a(\xi)$ and data points are more dispersed if we increase $\gamma$ (Fig.\ \ref{fig:scat}(b)).
For example, Fig.\ \ref{fig:scat}(d) shows a double logarithmic plot of $a_c$ against $\gamma$,
where all data collapse onto a linear scaling, $a_c\simeq A_c\gamma$, with $A_c=0.76\pm0.002$.
We also find other scaling relations, $a_v\simeq 0$, $b_c\simeq B_c\gamma$, $b_v\simeq B_v\gamma$, $v_c\simeq V_c\gamma$, and $v_v\simeq V_v\gamma$
with $B_c=0.24\pm0.002$, $B_v=1.80\pm0.001$, $V_c=0.32\pm0.01$, and $V_v=4.41\pm0.06$, respectively, for $\gamma<1$ (see the ESI \dag),
so that all parameters characterizing the mean values and fluctuations are linearly scaled by $\gamma$.
Because $a_v\simeq0$ and $B_v\approx B_a$($\simeq1.9$ for small and large particles), virtual contacts almost behave affine in average, except for their huge fluctuations ($V_v\gg V_c$).
In contrast to (CC) and (VV), the data of $\xi'$ in (VC) and (CV) are concentrated in narrow regions (between the axes and the dashed lines in Fig.\ \ref{fig:scat}(c)),
whereas $f_a(\xi)$ linearly increases with $\xi$ in (VC) and there is no data of $f_a(\xi)$ in (CV),\ i.e.\ the affine deformation gives closing contacts only.

\begin{figure}[h]
\centering
\includegraphics[width=8.5cm]{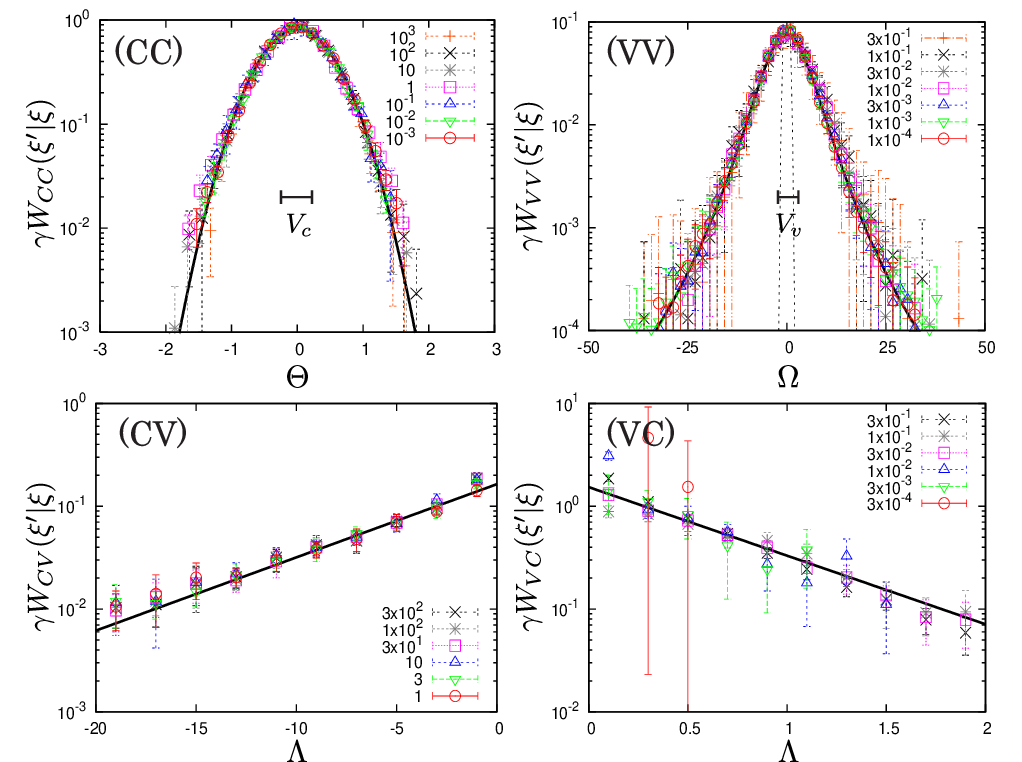}
\caption{(Color online)
Semi-logarithmic plots of the CPDs, where we fix $\xi=1.6$ (CC), $0.2$ (CV), and $-0.2$ (VC), respectively, while we average $W_{VV}(\xi'|\xi)$ over $-20\le\xi\le0$.
The different symbols represent $\gamma$, as given in the insets, and the solid lines are given by Eqs.\ (\ref{eq:cdn1})-(\ref{eq:cdn3})
(note the different horizontal axis scales). The dotted line in (VV) is a Gaussian distribution function with the width, $V_c$.
\label{fig:cpd}}
\end{figure}

We then determine the CPDs for non-affine deformations as the distributions of scaled overlaps, $\xi'$, around their mean values, $f_n(\xi)$.
Figure \ref{fig:cpd}(a) shows the CPDs in (CC), where all results with a wide range of $\gamma$ are symmetric around $f_c(\xi)$
and collapse if we multiply $W_{CC}(\xi'|\xi)$ and $\xi'-f_c(\xi)$ by $\gamma$ and $1/\gamma$, respectively.
The solid line is a Gaussian distribution function,
\begin{equation}
\gamma W_{CC}(\xi'|\xi)=\frac{1}{\sqrt{2\pi V_c^2}}\hspace{1mm}e^{-\Theta^2/2V_c^2}~,
\label{eq:cdn1}
\end{equation}
with $\Theta\equiv[\xi'-f_c(\xi)]/\gamma$.
Figure \ref{fig:cpd}(b) displays the CPDs in (VV), where all results are also symmetric around $f_v(\xi)$ and collapse as well, after the same scaling as for (CC).
The solid line is here a stable distribution function \cite{sdistribution},
\begin{equation}
\gamma W_{VV}(\xi'|\xi)=\frac{1}{2\pi}\int_{-\infty}^\infty
e^{-\left(\kappa|V_v z|^\lambda+i\Omega z\right)}dz~,
\label{eq:cdn2}
\end{equation}
with $\Omega\equiv[\xi'-f_v(\xi)]/\gamma$, where $z$ is a dimensionless wave number,
and the fitting parameters are given by $\lambda=1.65$ and $\kappa=0.62$, respectively,\ i.e.\ the CPD in (VV) is nearly a Holtsmark distributions ($\lambda=3/2$ and $\kappa>0$).
Figures \ref{fig:cpd}(c) and (d) show the CPDs in (CV) and (VC) approximated by exponential distributions,
\begin{eqnarray}
\gamma W_{CV}(\xi'|\xi) &=& \left\{1-I_{CC}(\xi)\right\}\frac{e^{ \Lambda/q_v}}{q_v}~, \label{eq:cdn4}\\
\gamma W_{VC}(\xi'|\xi) &=& \left\{1-I_{VV}(\xi)\right\}\frac{e^{-\Lambda/q_c}}{q_c}~, \label{eq:cdn3}
\end{eqnarray}
respectively, where $\Lambda\equiv\xi'/\gamma$ and the dimensionless lengths are given by $q_v=6.10$ and $q_c=0.65$ ($q_v\gg q_c$), respectively
\footnote[7]{
The meaning of $q_v$ is that $\gamma q_v$ represents a typical length of interparticle gaps which are generated by opening contacts.
Similarly, new contacts have a typical overlap $\sim\gamma q_c$.
For example, $\gamma q_v\simeq0.061$ and $\gamma q_c\simeq0.0065$ for $\gamma=0.01$ in our length scale.
}.
In curly brackets on the right hand sides, $I_{CC}(\xi)\equiv\frac{1}{2}\mathrm{erfc}\left[-\frac{f_c(\xi)}{\sqrt{2}v_c}\right]$
and $I_{VV}(\xi)\equiv\int_{-\infty}^0 W_{VV}(\xi'|\xi)d\xi'$ are the cumulative distribution functions of the CPDs in (CC) and (VV), respectively,
which are required to satisfy the normalization conditions
\footnote[8]{The normalization conditions are $\int_{-\infty}^0 W_{VV}d\xi'+\int_0^\infty W_{VC}d\xi'=\int_{-\infty}^0 W_{CV}d\xi'+\int_0^\infty W_{CC}d\xi'=1$,
for previously virtual contacts and contacts, respectively.}
and well describe the dependence of the CPDs on $\xi$ (see the ESI \dag).
In addition, if $\gamma=0$, $W_{CC}=W_{VV}=\delta(\xi-\xi')$ and $W_{CV}=W_{VC}=0$
\footnote[9]{We used
$W_{VV}=(2\pi)^{-1}\int e^{-\left[\kappa|\gamma V_v z|^\lambda+i(\xi'-f_v)z\right]}dz=(2\pi)^{-1}\int e^{i(\xi-\xi')z}dz\rightarrow\delta(\xi-\xi')$
and $e^{-1/\gamma}/\gamma\rightarrow0$ for $\gamma\rightarrow0$.}
so that the Chapman-Kolmogorov equation (\ref{eq:conditional}) does not change the PDF without deformations.

\begin{figure}[h]
\centering
\includegraphics[width=8.5cm]{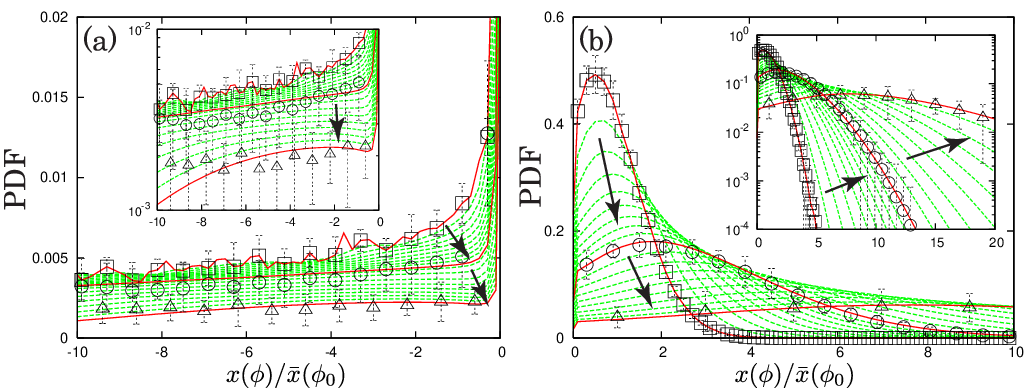}
\caption{(Color online)
Numerical solutions of the Master equation (the solid and dotted lines) under compression, where (a) and (b) display the PDFs of negative and positive overlaps, respectively.
The solutions develop in the directions indicated by the arrows.
The open squares, circles, and triangles are the PDFs obtained from MD simulations with $\phi-\phi_J=4\times10^{-3}$, $1.2\times10^{-2}$, and $4\times10^{-2}$, respectively.
The insets show the semi-logarithmic plots. Overlaps are scaled by the averaged overlap at $\phi_0-\phi_J=4\times10^{-3}$.
\label{fig:mrkv}}
\end{figure}

Now, we restrict $\delta\phi$ to quite small values compared to $\phi-\phi_J$
and define an infinitesimal scaled strain step as $\delta\gamma\equiv\delta\phi/(\phi-\phi_J)\ll1$.
Introducing a transition rate as $T(\xi'|\xi)=\lim_{\delta\gamma\rightarrow0}W(\xi'|\xi)/\delta\gamma$,
we rewrite the Chapman-Kolmogorov equation (\ref{eq:conditional}) as a Master equation \cite{vanKampen},
\begin{equation}
\frac{\partial}{\partial\gamma}P_\phi(\xi')=\int_{-\infty}^\infty
\left[T(\xi'|\xi)P_\phi(\xi)-T(\xi|\xi')P_\phi(\xi')\right]d\xi~,
\label{eq:master}
\end{equation}
where we use the CPDs, Eqs.\ (\ref{eq:cdn1})-(\ref{eq:cdn3}), for the transition rates.
Figures \ref{fig:mrkv}(a) and (b) display the numerical solutions of the Master equation under \emph{incremental} compression steps,
where the increment of area fraction is fixed to $\delta\phi=10^{-5}$ so that $\delta\gamma\le2.5\times10^{-3}$ throughout the numerical integrations.
Here, the initial condition is given by the PDF obtained through MD simulations with the distance from jamming, $\phi_0-\phi_J=4\times10^{-3}$.
The overlaps are scaled by the averaged overlap at the initial state, $\bar{x}(\phi_0)$.
Good agreements between the solutions (red solid lines) and MD simulations (open symbols) are established for small $\delta\gamma$ even in the tails of the PDFs (the inset in Fig.\ \ref{fig:mrkv}).
In addition, the Master equation reproduces discontinuous jumps of the PDFs around zero-overlap as observed in Fig.\ \ref{fig:delaunay}(b).
We also confirmed that numerical solutions starting from different initial conditions,\ e.g.\ a step function and a Gaussian distribution (not consistent with mechanical stability),
converge to a unique solution with discontinuous jumps around zero (see the ESI \dag).

In additional MD simulations of \emph{decompression} tests with the increment of area fraction, $\delta\phi<0$,
we find that the mean values and CPDs are given by just replacing the scaling parameter, $\gamma$, with $-\gamma$
in Eqs.\ (\ref{eq:mzt})-(\ref{eq:cdn3}), which does not change the form of the Master equation (\ref{eq:master}).
Therefore, the linear scalings of the coefficients for non-affine deformations, $a_n$, $b_n$, and $v_n$, are maintained under decompression,
and the functional forms of the CPDs are the same for both compression and decompression (see the ESI \dag).
However, note that the scattered data under compression and decompression are not symmetric with respect to the diagonal line, $\xi'=\xi$.
Thus, the transition rates for decompressions are $T_{\delta\gamma<0}(\xi'|\xi)\neq T_{\delta\gamma>0}(\xi|\xi')$,
which leads to \emph{irreversible} responses of soft particle packings under quasi-static cyclic (de)compressions.

In summary, we provide, for the first time, a Master equation for the PDFs of forces in soft particle packings under quasi-static (de)compressions,
where not only the changes of contacts and virtual contacts, but also their mutual exchange,\ i.e.\ opening and closing contacts, are included in the transition rates for the Master equation.
The transition rates (or the CPDs of the generalized overlaps) are symmetric around mean values with finite widths,
where both the mean and fluctuations are well characterized by a single scaling parameter,
$\gamma=\delta\phi/(\phi-\phi_J)$, quantifying the degree of non-affine deformations.
We confirm that shapes of the CPDs and linear scalings for the mean and fluctuations are the same for compression and decompression.
The Master equation can predict the incremental evolution of the PDFs, including discontinuous jumps around zero,
that is, the multi-particle system is reduced to a single-contact picture,\ i.e.\ a mean-field like description.

The CPDs show by themselves important properties:
Contacts respond in a non-affine way, especially near jamming \cite{rs1},
as quantified by the scaling,\ e.g.\ $a_c\sim\gamma=\delta\phi/\left(\phi-\phi_J\right)$.
Astonishingly, their fluctuations obey Gaussian statistics, indicating uncorrelated stochastic evolution of forces \cite{ft3}.
In contrast, the nearly Holtsmark distributions feature much broader tails for virtual contacts that deform affinely in average.
Indicating much larger changes of interparticle gaps, this implies a strongly correlated stochastic evolution over a wide range of length-scales.
The probabilities for opening and closing contacts are exponentially decaying with distance from zero
(i.e.\ $e^{-|\Lambda|/q_v}$ and $e^{-|\Lambda|/q_c}$ in Eqs.\ (\ref{eq:cdn4}) and (\ref{eq:cdn3}), respectively),
and cause the discontinuous jumps in the PDFs, since opening contacts are free to open widely
whereas closing contacts are affected by repulsion,\ i.e.\ $q_v\gg q_c$.
Because both the Gaussian and Holtsmark distributions are members of the stable distribution family,
fluctuations of contacts and virtual contacts in soft particles should obey the generalized central limit theorem \cite{sdistribution},
which has consequences for the statistical description of disordered systems in general.
The strong deviation from an affine approximation \cite{rs1} for contacts and the enormous fluctuations of overlaps \cite{ft2} for virtual contacts,
as well as the probabilities for opening and closing of contacts, are all proportional to the scaled strain step, $\gamma$.

Clearly, there is the need of further studies on the physical origin of the statistics of overlaps described above.
The functional forms of the CPDs can give very interesting insights into the micro-mechanics of soft particles,\ e.g.\ stochastic processes of overlaps in force-chain networks.
Now, analytic solutions or asymptotic solutions of the Master equation are important next steps towards the understanding of the functional forms of the PDFs.
The Master equation also poses a new challenge; it requires the increment $\delta\phi$ to be much smaller than $\phi-\phi_J$,\ i.e. $\gamma\ll1$.
Thus, strictly speaking, it can never reach $\phi_J$, and the result cannot be the PDF at $\phi_J$, albeit asymptotically.
This means that the jamming transition is a singular limit of the Master equation.

Finally, our analysis can be easily extended to three dimensions and be examined and validated by experiments,\
e.g.\ by photoelastic tests \cite{an0} or oedometer test of sands \cite{wangao,imole}.
The extension to other cases is also straightforward,\ e.g.\ the solutions under shear can be obtained
if we apply our results for (de)compressions to each principal direction (in preparation).

We thank M. Sperl, L.E. Silbert, B.P. Tighe, H. Hayakawa, S. Yukawa, T. Hatano, H. Yoshino, K. Kanazawa for fruitful discussions.
This work was financially supported by the NWO-STW VICI grant 10828 and a part of numerical
computation has been carried out at the Yukawa Institute Computer Facility, Kyoto, Japan.
\footnotesize{
\bibliography{master}

\providecommand*{\mcitethebibliography}{\thebibliography}
\csname @ifundefined\endcsname{endmcitethebibliography}
{\let\endmcitethebibliography\endthebibliography}{}
\begin{mcitethebibliography}{28}
\providecommand*{\natexlab}[1]{#1}
\providecommand*{\mciteSetBstSublistMode}[1]{}
\providecommand*{\mciteSetBstMaxWidthForm}[2]{}
\providecommand*{\mciteBstWouldAddEndPuncttrue}
  {\def\EndOfBibitem{\unskip.}}
\providecommand*{\mciteBstWouldAddEndPunctfalse}
  {\let\EndOfBibitem\relax}
\providecommand*{\mciteSetBstMidEndSepPunct}[3]{}
\providecommand*{\mciteSetBstSublistLabelBeginEnd}[3]{}
\providecommand*{\EndOfBibitem}{}
\mciteSetBstSublistMode{f}
\mciteSetBstMaxWidthForm{subitem}
{(\emph{\alph{mcitesubitemcount}})}
\mciteSetBstSublistLabelBeginEnd{\mcitemaxwidthsubitemform\space}
{\relax}{\relax}

\bibitem[Lemaitre and Chaboche(1990)]{lemaitre}
J.~Lemaitre and J.-L. Chaboche, \emph{Mechanics of Solid Materials}, Cambridge
  University Press, Cambridge, UK, 1990\relax
\mciteBstWouldAddEndPuncttrue
\mciteSetBstMidEndSepPunct{\mcitedefaultmidpunct}
{\mcitedefaultendpunct}{\mcitedefaultseppunct}\relax
\EndOfBibitem
\bibitem[Majmudar and Behringer(2005)]{an0}
T.~S. Majmudar and R.~P. Behringer, \emph{Nature}, 2005, \textbf{435},
  1079\relax
\mciteBstWouldAddEndPuncttrue
\mciteSetBstMidEndSepPunct{\mcitedefaultmidpunct}
{\mcitedefaultendpunct}{\mcitedefaultseppunct}\relax
\EndOfBibitem
\bibitem[Majmudar \emph{et~al.}(2007)Majmudar, Sperl, Luding, and
  Behringer]{gn2}
T.~S. Majmudar, M.~Sperl, S.~Luding and R.~P. Behringer, \emph{Phys.\ Rev.\
  Lett.}, 2007, \textbf{98}, 058001\relax
\mciteBstWouldAddEndPuncttrue
\mciteSetBstMidEndSepPunct{\mcitedefaultmidpunct}
{\mcitedefaultendpunct}{\mcitedefaultseppunct}\relax
\EndOfBibitem
\bibitem[van Deen \emph{et~al.}(2014)van Deen, Simon, Zeravcic, D.-Bohy, Tighe,
  and van Hecke]{merlijn}
M.~S. van Deen, J.~Simon, Z.~Zeravcic, S.~D.-Bohy, B.~P. Tighe and M.~van
  Hecke, \emph{Phys.\ Rev.\ E}, 2014, \textbf{90}, 020202(R)\relax
\mciteBstWouldAddEndPuncttrue
\mciteSetBstMidEndSepPunct{\mcitedefaultmidpunct}
{\mcitedefaultendpunct}{\mcitedefaultseppunct}\relax
\EndOfBibitem
\bibitem[Henkes and Chakraborty(2009)]{ft4}
S.~Henkes and B.~Chakraborty, \emph{Phys.\ Rev.\ E}, 2009, \textbf{79},
  061301\relax
\mciteBstWouldAddEndPuncttrue
\mciteSetBstMidEndSepPunct{\mcitedefaultmidpunct}
{\mcitedefaultendpunct}{\mcitedefaultseppunct}\relax
\EndOfBibitem
\bibitem[Snoeijer \emph{et~al.}(2004)Snoeijer, Vlugt, van Hecke, and van
  Saarloos]{en4}
J.~H. Snoeijer, T.~J.~H. Vlugt, M.~van Hecke and W.~van Saarloos, \emph{Phys.\
  Rev.\ Lett.}, 2004, \textbf{92}, 054302\relax
\mciteBstWouldAddEndPuncttrue
\mciteSetBstMidEndSepPunct{\mcitedefaultmidpunct}
{\mcitedefaultendpunct}{\mcitedefaultseppunct}\relax
\EndOfBibitem
\bibitem[Tighe \emph{et~al.}(2008)Tighe, van Eerd, and Vlugt]{en7}
B.~P. Tighe, A.~R.~T. van Eerd and T.~J.~H. Vlugt, \emph{Phys.\ Rev.\ Lett.},
  2008, \textbf{100}, 238001\relax
\mciteBstWouldAddEndPuncttrue
\mciteSetBstMidEndSepPunct{\mcitedefaultmidpunct}
{\mcitedefaultendpunct}{\mcitedefaultseppunct}\relax
\EndOfBibitem
\bibitem[Metzger(2004)]{ft0}
P.~T. Metzger, \emph{Phys.\ Rev.\ E}, 2004, \textbf{70}, 051303\relax
\mciteBstWouldAddEndPuncttrue
\mciteSetBstMidEndSepPunct{\mcitedefaultmidpunct}
{\mcitedefaultendpunct}{\mcitedefaultseppunct}\relax
\EndOfBibitem
\bibitem[Henkes \emph{et~al.}(2007)Henkes, O'Hern, and Chakraborty]{ft3}
S.~Henkes, C.~S. O'Hern and B.~Chakraborty, \emph{Phys.\ Rev.\ Lett.}, 2007,
  \textbf{99}, 038002\relax
\mciteBstWouldAddEndPuncttrue
\mciteSetBstMidEndSepPunct{\mcitedefaultmidpunct}
{\mcitedefaultendpunct}{\mcitedefaultseppunct}\relax
\EndOfBibitem
\bibitem[Corwin \emph{et~al.}(2005)Corwin, Jaeger, and Nagel]{ps1}
E.~I. Corwin, H.~M. Jaeger and S.~R. Nagel, \emph{Nature}, 2005, \textbf{435},
  1075\relax
\mciteBstWouldAddEndPuncttrue
\mciteSetBstMidEndSepPunct{\mcitedefaultmidpunct}
{\mcitedefaultendpunct}{\mcitedefaultseppunct}\relax
\EndOfBibitem
\bibitem[Desmond \emph{et~al.}(2013)Desmond, Young, Chen, and Weeks]{ps2}
K.~W. Desmond, P.~J. Young, D.~Chen and E.~R. Weeks, \emph{Soft Matter}, 2013,
  \textbf{9}, 3424\relax
\mciteBstWouldAddEndPuncttrue
\mciteSetBstMidEndSepPunct{\mcitedefaultmidpunct}
{\mcitedefaultendpunct}{\mcitedefaultseppunct}\relax
\EndOfBibitem
\bibitem[Silbert \emph{et~al.}(2002)Silbert, Grest, and Landry]{pd3}
L.~E. Silbert, G.~S. Grest and J.~W. Landry, \emph{Phys.\ Rev.\ E}, 2002,
  \textbf{66}, 061303\relax
\mciteBstWouldAddEndPuncttrue
\mciteSetBstMidEndSepPunct{\mcitedefaultmidpunct}
{\mcitedefaultendpunct}{\mcitedefaultseppunct}\relax
\EndOfBibitem
\bibitem[Landry \emph{et~al.}(2003)Landry, Grest, Silbert, and Plimpton]{pd4}
J.~W. Landry, G.~S. Grest, L.~E. Silbert and S.~J. Plimpton, \emph{Phys.\ Rev.\
  E}, 2003, \textbf{67}, 041303\relax
\mciteBstWouldAddEndPuncttrue
\mciteSetBstMidEndSepPunct{\mcitedefaultmidpunct}
{\mcitedefaultendpunct}{\mcitedefaultseppunct}\relax
\EndOfBibitem
\bibitem[M{\" u}ller \emph{et~al.}(2010)M{\" u}ller, Luding, and P{\"
  o}schel]{gu4}
M.-K. M{\" u}ller, S.~Luding and T.~P{\" o}schel, \emph{Chem.\ Phys.}, 2010,
  \textbf{375}, 600\relax
\mciteBstWouldAddEndPuncttrue
\mciteSetBstMidEndSepPunct{\mcitedefaultmidpunct}
{\mcitedefaultendpunct}{\mcitedefaultseppunct}\relax
\EndOfBibitem
\bibitem[Radjai \emph{et~al.}(1996)Radjai, Jean, Moreau, and Roux]{pd0}
F.~Radjai, M.~Jean, J.-J. Moreau and S.~Roux, \emph{Phys.\ Rev.\ Lett.}, 1996,
  \textbf{77}, 274\relax
\mciteBstWouldAddEndPuncttrue
\mciteSetBstMidEndSepPunct{\mcitedefaultmidpunct}
{\mcitedefaultendpunct}{\mcitedefaultseppunct}\relax
\EndOfBibitem
\bibitem[van Eerd \emph{et~al.}(2007)van Eerd, Ellenbroek, van Hecke, Snoeijer,
  and Vlugt]{gu5}
A.~R.~T. van Eerd, W.~G. Ellenbroek, M.~van Hecke, J.~H. Snoeijer and T.~J.~H.
  Vlugt, \emph{Phys.\ Rev.\ E}, 2007, \textbf{75}, 060302(R)\relax
\mciteBstWouldAddEndPuncttrue
\mciteSetBstMidEndSepPunct{\mcitedefaultmidpunct}
{\mcitedefaultendpunct}{\mcitedefaultseppunct}\relax
\EndOfBibitem
\bibitem[Wyart(2012)]{wyart0}
M.~Wyart, \emph{Phys.\ Rev.\ Lett.}, 2012, \textbf{109}, 125502\relax
\mciteBstWouldAddEndPuncttrue
\mciteSetBstMidEndSepPunct{\mcitedefaultmidpunct}
{\mcitedefaultendpunct}{\mcitedefaultseppunct}\relax
\EndOfBibitem
\bibitem[Lerner \emph{et~al.}(2013)Lerner, D{\" u}ring, and Wyart]{wyart1}
E.~Lerner, G.~D{\" u}ring and M.~Wyart, \emph{Soft Matter}, 2013, \textbf{9},
  8252\relax
\mciteBstWouldAddEndPuncttrue
\mciteSetBstMidEndSepPunct{\mcitedefaultmidpunct}
{\mcitedefaultendpunct}{\mcitedefaultseppunct}\relax
\EndOfBibitem
\bibitem[Charbonneau \emph{et~al.}(2012)Charbonneau, Corwin, Parisi, and
  Zamponi]{wyart2}
P.~Charbonneau, E.~I. Corwin, G.~Parisi and F.~Zamponi, \emph{Phys.\ Rev.\
  Lett.}, 2012, \textbf{109}, 205501\relax
\mciteBstWouldAddEndPuncttrue
\mciteSetBstMidEndSepPunct{\mcitedefaultmidpunct}
{\mcitedefaultendpunct}{\mcitedefaultseppunct}\relax
\EndOfBibitem
\bibitem[Berthier \emph{et~al.}(2011)Berthier, Jacquin, and Zamponi]{th1}
L.~Berthier, H.~Jacquin and F.~Zamponi, \emph{Phys.\ Rev.\ E}, 2011,
  \textbf{84}, 051103\relax
\mciteBstWouldAddEndPuncttrue
\mciteSetBstMidEndSepPunct{\mcitedefaultmidpunct}
{\mcitedefaultendpunct}{\mcitedefaultseppunct}\relax
\EndOfBibitem
\bibitem[Ellenbroek \emph{et~al.}(2009)Ellenbroek, van Hecke, and van
  Saarloos]{rs1}
W.~G. Ellenbroek, M.~van Hecke and W.~van Saarloos, \emph{Phys.\ Rev.\ E},
  2009, \textbf{80}, 061307\relax
\mciteBstWouldAddEndPuncttrue
\mciteSetBstMidEndSepPunct{\mcitedefaultmidpunct}
{\mcitedefaultendpunct}{\mcitedefaultseppunct}\relax
\EndOfBibitem
\bibitem[van Hecke(2010)]{gn3}
M.~van Hecke, \emph{J.\ Phys.:\ Condens.\ Matter}, 2010, \textbf{22},
  033101\relax
\mciteBstWouldAddEndPuncttrue
\mciteSetBstMidEndSepPunct{\mcitedefaultmidpunct}
{\mcitedefaultendpunct}{\mcitedefaultseppunct}\relax
\EndOfBibitem
\bibitem[Chaudhuri \emph{et~al.}(2010)Chaudhuri, Berthier, and Sastry]{sastry}
P.~Chaudhuri, L.~Berthier and S.~Sastry, \emph{Phys.\ Rev.\ Lett.}, 2010,
  \textbf{104}, 165701\relax
\mciteBstWouldAddEndPuncttrue
\mciteSetBstMidEndSepPunct{\mcitedefaultmidpunct}
{\mcitedefaultendpunct}{\mcitedefaultseppunct}\relax
\EndOfBibitem
\bibitem[van Kampen(2007)]{vanKampen}
N.~G. van Kampen, \emph{Stochastic Processes in Physics and Chemistry, 3rd
  edition}, Elsevier B. V. Amsterdam, The Netherlands, 2007\relax
\mciteBstWouldAddEndPuncttrue
\mciteSetBstMidEndSepPunct{\mcitedefaultmidpunct}
{\mcitedefaultendpunct}{\mcitedefaultseppunct}\relax
\EndOfBibitem
\bibitem[Voit(2005)]{sdistribution}
J.~Voit, \emph{The Statistical Mechanics of Financial Markets, 3rd Edition},
  Springer-Verlag, Berlin, 2005\relax
\mciteBstWouldAddEndPuncttrue
\mciteSetBstMidEndSepPunct{\mcitedefaultmidpunct}
{\mcitedefaultendpunct}{\mcitedefaultseppunct}\relax
\EndOfBibitem
\bibitem[Henkes and Chakraborty(2005)]{ft2}
S.~Henkes and B.~Chakraborty, \emph{Phys.\ Rev.\ Lett.}, 2005, \textbf{95},
  198002\relax
\mciteBstWouldAddEndPuncttrue
\mciteSetBstMidEndSepPunct{\mcitedefaultmidpunct}
{\mcitedefaultendpunct}{\mcitedefaultseppunct}\relax
\EndOfBibitem
\bibitem[Wang and Gao(2014)]{wangao}
Y.-H. Wang and Y.~Gao, \emph{Granular Matter}, 2014, \textbf{16}, 55\relax
\mciteBstWouldAddEndPuncttrue
\mciteSetBstMidEndSepPunct{\mcitedefaultmidpunct}
{\mcitedefaultendpunct}{\mcitedefaultseppunct}\relax
\EndOfBibitem
\bibitem[Imole \emph{et~al.}(2014)Imole, Wojtkowski, Magnanimo, and
  Luding]{imole}
O.~I. Imole, M.~Wojtkowski, V.~Magnanimo and S.~Luding, \emph{Phys.\ Rev.\ E},
  2014, \textbf{89}, 042210\relax
\mciteBstWouldAddEndPuncttrue
\mciteSetBstMidEndSepPunct{\mcitedefaultmidpunct}
{\mcitedefaultendpunct}{\mcitedefaultseppunct}\relax
\EndOfBibitem
\end{mcitethebibliography}
\bibliographystyle{rsc}}
\end{document}